\documentclass[sigconf,nonacm]{acmart}

\usepackage{hyperref}

\usepackage{todonotes}

\usepackage{enumitem}  

\def\eg{e.g.,}         
\def\ie{i.e.,}         
\def\aka{aka}     

\usepackage{listings}

\definecolor{UVLkeyword}{HTML}{a586c0}       
\definecolor{UVLattribute}{HTML}{4e94ce}     
\definecolor{UVLstring}{HTML}{ce723b}        
\definecolor{UVLcomment}{HTML}{529955}       
\definecolor{UVLbrackets}{HTML}{ffd700}      
\definecolor{UVLnumbers}{HTML}{b5cea8}       
\definecolor{UVLconstraints}{HTML}{d0a343}   

\lstdefinelanguage{uvl}{ 
    basicstyle=\ttfamily\scriptsize, 
	keywords = {alternative, or, optional, mandatory, features, constraints, constraint, cardinality, Real, Boolean, Integer, String, imports, as, include, true, false, attributes, xor, namespace},
	morecomment=[l]{//},
	commentstyle=\color{UVLcomment},
    keywordstyle=\ttfamily\color{UVLkeyword},
    tabsize=2,
    columns=fullflexible,
    keywordstyle = [2]{\color{UVLattribute}}, 
    morekeywords = [2]{abstract, default, Price, size, vegan, Volume, Alcohol, doc, unit, desc, medium, big, ABV, value, currency, Units},
    keywordstyle = [3]{\color{UVLconstraints}}, 
    morekeywords = [3]{sum, len, avg, ceil, floor, forall, all, disj, index, children, max},
    moredelim = [s][\color{UVLstring}]{''},
    frame = lines,
    xleftmargin = 0pt,
	framexleftmargin = 0pt,
	framesep = 0pt,
    numbers = left,
	numberstyle = \tiny,
	numbersep = 2pt,
    breaklines = true,        
    extendedchars=true,       
    literate={€}{{\EUR}}1 %
      {\ \ \ \ \ \ \ \ \ \ \ \ \ \ \ \ \ \ \ \ }{{\ \ \ \ \ \ \ \ \ \ \ \ }}{16} 
      {\ \ \ \ \ \ \ \ \ \ \ \ \ \ \ \ }{{\ \ \ \ \ \ \ \ }}{12} 
      {\ \ \ \ \ \ \ \ \ \ \ \ }{{\ \ \ \ \ \ \ \ }}8 
      {\ \ \ \ \ \ \ \ }{{\ \ \ \ }}4 
      {\ \ \ \ }{{\ \ }}2             
}

\lstnewenvironment{uvlfile}[1][]
{
    \lstset{belowskip=-0.4cm,language=uvl, #1}
}
{}

\newcommand{\UVLKeyword}[1]{{{\textcolor{UVLkeyword}{\texttt{#1}}}}}
\newcommand{\feature}[1]{{{\texttt{#1}}}}

\definecolor{TexGreen}{HTML}{338596}
\definecolor{TexBlue}{HTML}{0B0BFF}
\definecolor{JinjaBrown}{HTML}{9B5D37}
\definecolor{JinjaBlue}{HTML}{2054A2}

\lstdefinelanguage{jinja}{
	language = {},
	tabsize = 2,
	frame = lines,
	xleftmargin = 0pt,
	framexleftmargin = 0pt,
	framesep = 0pt,
	numbers = left,
	numberstyle = \tiny,
	numbersep = 2pt,
	breaklines = true,
	showstringspaces = false,
    alsoletter={\\},
    moredelim = [s][\bfseries]{\{\{}{\}\}},
    moredelim = [s][\bfseries]{\{\%}{\%\}},
	basicstyle = {\scriptsize \ttfamily},
    keywordstyle = [2]{\color{TexBlue}}, 
    morekeywords = [2]{\\begin,\\end,\\addplot,\\addlegendentry,\\textwidth},
    keywordstyle = [3]{\color{TexGreen}}, 
    morekeywords = [3]{tikzpicture,axis},
    keywordstyle = [5]{\color{JinjaBlue}\pmb}, 
    morekeywords = [5]{for,in,endfor,if,endif},
    literate=
      {\ \ \ \ \ \ \ \ \ \ \ \ \ \ \ \ \ \ \ \ }{{\ \ \ \ \ \ \ \ \ \ \ \ }}{16} 
      {\ \ \ \ \ \ \ \ \ \ \ \ \ \ \ \ }{{\ \ \ \ \ \ \ \ }}{12} 
      {\ \ \ \ \ \ \ \ \ \ \ \ }{{\ \ \ \ \ \ \ \ }}8 
      {\ \ \ \ \ \ \ \ }{{\ \ \ \ }}4 
      {\ \ \ \ }{{\ \ }}2             
}

\lstnewenvironment{jinjatemplate}[1][]
{
    \lstset{belowskip=-0.4cm,language=jinja, #1}
}
{}
\AtBeginDocument{%
  \providecommand\BibTeX{{%
    \normalfont B\kern-0.5em{\scshape i\kern-0.25em b}\kern-0.8em\TeX}}}

\begin{document}

\title{Capturing and Exploiting Design Pattern Variability in Mobile Application Generation}

\settopmatter{authorsperrow=2}
\author{Ramón Peralta}
\affiliation{%
  \institution{Universidad de San Jorge}
  \city{Zaragoza}
  \country{Spain}
}
\email{anperalta@gmail.com}
\author{Jose-Miguel Horcas}
\orcid{0000-0002-7771-0575}
\affiliation{%
  \institution{ITIS Software, Universidad de Málaga}
  \city{Málaga}
  \country{Spain}
}
\email{horcas@uma.es}
\renewcommand{\shortauthors}{R. Peralta et al.}

\begin{abstract}
The increasing reliance on automatic code generation in mobile application development often leads to code that neglects fundamental design principles and architectural quality. In this work, we address this challenge by capturing and exploiting the inherent variability of software design patterns to systematically generate customizable and well-structured mobile applications. We propose the use of the Universal Variability Language (UVL) to explicitly model the structural and behavioral variation points of common design patterns, such as Singleton, Strategy, Observer, Adapter, and Factory Method. These models are integrated with reusable Jinja templates, enabling code generation in Swift. Our approach leverages Software Product Line (SPL) engineering principles, treating design patterns as configurable assets within a product line and supporting automated generation of custom design patterns. We also analyze the configuration space of the modeled patterns, offering insights into their variability complexity. By formalizing design pattern variability and embedding it into the generation process, our work bridges model-driven engineering with practical mobile development, promoting the production of maintainable, reusable, and architecturally sound applications.
\end{abstract}



\maketitle

\section{Introduction}
\label{sec:Intro}
Mobile applications today are developed under increasing pressure to deliver rapidly, support multiple platforms, and adapt to evolving user expectations. To meet these demands, many developers turn to automatic code generation tools, especially those powered by \emph{large language models} (LLMs)~\cite{Juyong2024_SurveyLLMsCodeGeneration}. These tools can quickly produce code in languages such as Swift, Kotlin, or JavaScript. However, the generated code often lacks architectural quality and omits fundamental design principles~\cite{Mulla2024_ChoosingBestArchitecture}. In particular, the systematic application of well-established software design patterns (\ie\ key reusable solutions to recurring design problems) has become less explicit in this new paradigm~\cite{Nguyen2018_DeepLearningUIDesignPatterns}.

\emph{Design patterns}, as introduced by Gamma et al.~\cite{Gamma1995_DesignPatterns}, have long served as foundational building blocks for software design, facilitating code reuse, flexibility, and maintainability. In the context of mobile development, design patterns are particularly valuable. They help organize complex user interfaces~\cite{Punchoojit2017_UsabilityStudiesSLR}, manage application state~\cite{Zaina2022_GuidelinesUIDesignPatterns}, user interactions~\cite{DaSilva2022_MobileUIDesignPatternsSMS}, and support reusable business logic across platforms~\cite{Orlova2024_FlutterDesignPatterns}. Nevertheless, with increasing reliance on automatic code generation, developers may no longer recognize when or how to apply these patterns correctly~\cite{Orlova2024_FlutterDesignPatterns}.

Moreover, design patterns do not only provide reusable solutions at the code level; they also reflect architectural choices that vary depending on the needs of each application. This idea connects naturally with the principles of \emph{Software Product Line (SPL)} engineering~\cite{Pohl2005_SPLEngineering,Apel2013_FOSPL}, which focuses on developing families of related software products by managing their commonality and variability. In an SPL, software artifacts are designed to be reused across products, with variation points that allow for customization. \emph{Feature models}~\cite{Kang1990_FODA,Felfernig2024_FMsAIDriven} are commonly used to capture these options and guide the generation of concrete products. In the mobile development domain, where applications often need to be customized for different devices, brands, or user preferences, the combination of SPLs and design patterns provides a structured way to generate tailored applications from a shared architecture.

In SPL engineering, existing work often considers design patterns as tools to implement variability in software systems. For example, the \emph{State} or \emph{Strategy} patterns are commonly used in SPLs to enable the selection of alternative behaviors~\cite{Apel2013_FOSPL,Czarnecki2000_GenerativeProgramming}.  However, while patterns are often applied to handle variability, the variability internal to the patterns themselves has received comparatively little attention~\cite{Seidl2017_GenerativeSPLDesignPatterns}.

Before design patterns can be applied in a reusable and configurable way, we need to understand that they are not fixed templates. In practice, each design pattern has its own internal variability. For instance, the \emph{Observer} pattern can be instantiated with one or many observers, and the notification can be synchronous or asynchronous. Similarly, the \emph{Singleton} pattern has variants such as eager vs. lazy initialization, thread-safe vs. unsafe, or per-instance vs. global.
This internal variability is often implicit in design pattern catalogs and tutorials~\cite{Gamma1995_DesignPatterns,Freeman2004_HeadFirstDesignPatterns,Shvets2018_DiveIntoDesignPatterns}. As a result, developers applying these patterns manually (or through code generation) are left with an under-specified solution space. This makes automatic generation difficult to control, and pattern application prone to errors or misuse.

This article addresses two key problems: (1) how can we explicitly model the variability of commonly used design patterns?; and (2) how can we automatically generate Swift code from such models to produce correct and customizable pattern implementations for mobile applications?
To solve these problems, we propose the use of the \emph{Universal Variability Language} (UVL)~\cite{Benavides2025_UVL} to specify the variability of design patterns using feature modeling. UVL has become the de-facto standard for variability modeling in the SPL community in recent years~\cite{Sundermann2021_UVL}. UVL supports advanced constructs~\cite{Sundermann2023_UVLParserExtensions} such as attribute-based constraints, typed features beyond Boolean such as numeric and string features, and clonable elements (\aka\ feature cardinalities~\cite{Czarnecki2005_CardinalityBasedFM}), which are essential to accurately describe pattern variants. We then use a \emph{template-based
code generation}~\cite{Syriani2018_SMS_TemplateCodeGeneration} approach to implement each pattern, supporting variability directly in the code generation process~\cite{Horcas2025_UVengine}. In a template-based approach, a text-based programming language (\eg\ Swift, Kotlin, JavaScript, Python) is enriched with directives to automatically generate custom content.
Specifically, we leverage the \emph{Jinja}\footnote{\emph{Jinja}: \url{https://palletsprojects.com/p/jinja/}} template engine due to its language-agnostic nature, which allows its integration with various mobile programming languages, including Swift, Kotlin, and Java. Furthermore, Jinja supports all UVL extensions by using powerful directives to manage variability~\cite{Horcas2025_UVengine}. These directives facilitate the substitution and replacement of elements, the assignment of values, and the use of control structures such as \emph{if/elif/else}, \emph{for-loops}, \emph{macros}, and \emph{blocks}.
Our main contributions are as follows:
\begin{itemize}[noitemsep,nolistsep,leftmargin=\parindent]
    \item The explicit modeling of design pattern variability using UVL feature models, capturing both structural and behavioral variation points within each pattern.
    \item The definition of reusable Jinja templates for each pattern, enabling scalable and customizable code generation in Swift for mobile applications.
    \item The integration with existing SPL tools, such as \emph{UVengine}\footnote{\emph{UVengine}: \url{https://uvengine.github.io/}}~\cite{Horcas2025_UVengine}, to support automatic configuration and generation of pattern-based architectures tailored to user needs.
    \item The systematic analysis of the configuration space of each design pattern, providing insight into the variability complexity and supporting informed decision-making during product derivation.
\end{itemize}
By formalizing the variability inherent in design patterns, we lay the groundwork for tools that can assist developers (or even Artificial Intelligence assistants) in generating structured, reusable, and well-architected mobile applications. Our approach bridges the gap between model-driven engineering and practical code generation, ensuring that software quality and design best practices are not lost in the automation process.

The rest of the paper is organized as follows. Section~\ref{sec:Background} introduces the main concepts about SPL and UVL. Section~\ref{sec:Idea} presents our approach for modeling and mananing the variability of the design patterns, while Section~\ref{sec:Details} details the variability and implementation of each the design pattern. Section~\ref{sec:Evaluation} evaluates our approach. Section~\ref{sec:RelatedWork} discusess related work and Section~\ref{sec:Conclusions} concludes the paper.

\section{Background}
\label{sec:Background}
This section provides the foundational knowledge necessary to understand our work, focusing on the core principles of SPL engineering and the UVL language.

\subsection{Software Product Line Engineering}
\label{ssec:SPL}
SPL engineering is a paradigm for developing families of related software systems from a shared set of reusable assets in a planned and systematic manner~\cite{Pohl2005_SPLEngineering,Apel2013_FOSPL}. The central idea of SPLs is to balance the trade-off between the economies of scale achieved through mass production and the need for customization to meet specific market or customer demands. This is accomplished by explicitly managing two key concepts: 
\emph{commonality}, which refers to the features and assets shared by all products in the line, and 
\emph{variability}, which represents the points where products can differ. 
The SPL approach is typically organized into two main processes:
\begin{description}[noitemsep,nolistsep,leftmargin=\parindent]
    \item[Domain Engineering:] This phase focuses on the development and maintenance of reusable assets for the entire SPL. It includes activities such as domain analysis (identifying common and variable features), domain design (creating a common architecture), and domain implementation (developing reusable components).
    \item[Application Engineering:] This phase involves deriving specific products from the reusable assets created during domain engineering. It consists of selecting a valid combination of features from the variability model, which then guides the configuration and assembly of the reusable assets to generate a final product.
\end{description}

\subsection{The Universal Variability Language}
\label{ssec:UVL}
\begin{figure}[h!]
    \centering
    \vspace{-0.2cm}
    \includegraphics[width=\linewidth]{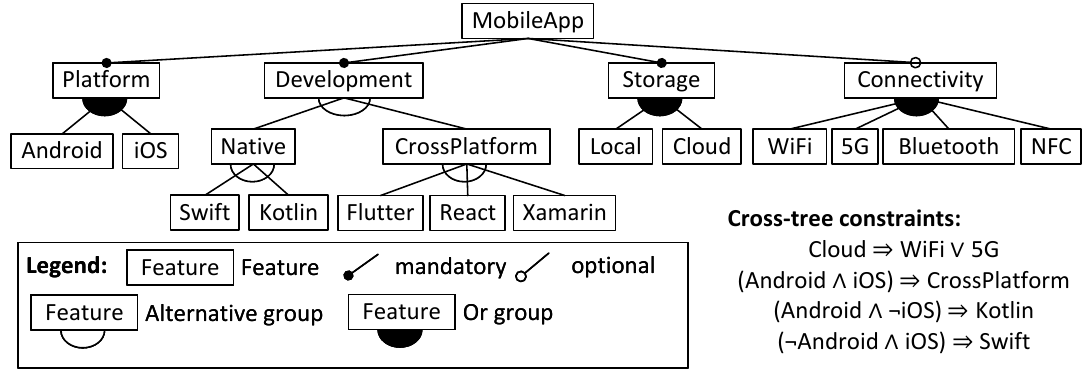}
    \vspace{-0.8cm}
    \caption{Diagram of the feature model of Listing~\ref{uvl:ExampleFM}.}
    \label{fig:ExampleFM}
    \vspace{-0.4cm}
\end{figure}

The SPL engineering process is driven by the \emph{variability model} (or \emph{feature model}), which specifies the commonality and variability of the entire SPL. Over the years, multiple languages and formats have been proposed to define feature models; however, in recent years, the UVL language has emerged as the de-facto standard within the SPL community~\cite{Benavides2025_UVL}.
Listing~\ref{uvl:ExampleFM} shows an example of a UVL model using its core Boolean level to represent the variability of a mobile application development process. Traditionally, feature models have been represented as feature diagrams, as illustrated in Figure~\ref{fig:ExampleFM}, but UVL offers a concise and expressive syntax for defining feature models.  

A UVL model consists of two main parts: the \UVLKeyword{features} section, which contains the feature tree, and the \UVLKeyword{constraints} section, which contains the cross-tree constraints. Features are organized in a tree-like hierarchy, definings relationships using keywords such as: \UVLKeyword{mandatory} to indicate that a child feature must be selected (\eg\, the \feature{Platform} feature); \UVLKeyword{optional} to indicate that a child feature may be selected (\eg\, the \feature{Connectivity} feature); \UVLKeyword{or} to indicate that at least one child feature must be selected in a group (\eg\ \feature{Platform} or \feature{Storage}); and \UVLKeyword{alternative} to indicate that exactly one child feature must be selected within a group (\eg\ \feature{Development}).
Cross-tree constraints are propositional formulas with logical operators: \texttt{!} (not), \texttt{\&} (and), \texttt{|} (or), \texttt{=>} (implies), \texttt{<=>} (equivalence), allowing further restrict valid feature selections (\eg\ \texttt{Cloud => WiFi | 5G}).

Furthermore, UVL also supports specifying \emph{feature attributes} to include additional information about the features by attaching key–value properties to features, and \emph{group cardinalities} \UVLKeyword{[a..b]} to specify the number of features that can be selected from a group. Other advanced variability constructs supported by UVL are \emph{typed features} (\UVLKeyword{Integer}, \UVLKeyword{Real}, or \UVLKeyword{String} features) that require assigning a concrete value during configuration; and \emph{feature cardinalities} (\aka\ clonable features), which allows creating multiple instances (clones) of a feature and configuring each clone and its subtree differently.

\begin{lstlisting}[language=uvl,caption={Example of feature model in UVL.}, label={uvl:ExampleFM}]
features
    MobileApp
        mandatory
            Platform
                or
                    Android
                    iOS
            Development
                alternative
                    Native
                        alternative
                            Swift
                            Kotlin
                    CrossPlatform
                        alternative
                            Flutter
                            React
                            Xamarin
            Storage
                or
                    Local
                    Cloud
        optional
            Connectivity
                or
                  WiFi
                  "5G"
                  Bluetooth
                  NFC
constraints
    Cloud => WiFi | "5G"
    (Android & iOS) => CrossPlatform
    (Android & !iOS) => Kotlin
    (!Android & iOS) => Swift
\end{lstlisting}

\section{Approach for Modeling and Exploiting Design Pattern Variability}
\label{sec:Idea}
Building upon the foundations of SPL engineering~\cite{Pohl2005_SPLEngineering} and Generative Programming~\cite{Czarnecki2000_GenerativeProgramming}, our work proposes a novel approach to address the challenge of creating well-architected and customizable mobile applications. The core idea is to treat design patterns themselves as reusable assets within an SPL, explicitly modeling and exploiting their inherent variability to systematically generate mobile application code. This shifts the focus from using design patterns as a static solution for implementing variability to treating them as a configurable, generative artifact.

\begin{figure}[t]
    \centering
    \includegraphics[width=\linewidth]{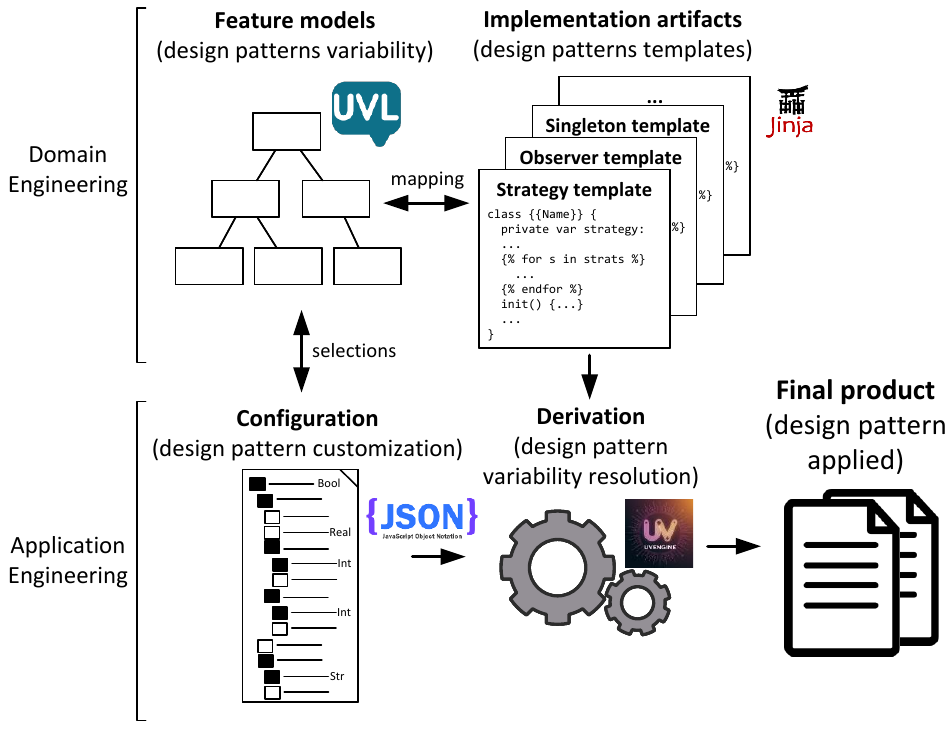}
    \vspace{-0.8cm}
    \caption{Our SPL approach to manage the variability of the design patterns.}
    \label{fig:Approach}
    \vspace{-0.4cm}
\end{figure}

This approach aligns directly with the two core phases of the SPL engineering process introduced in Section~\ref{sec:Background}, as illustrated in the conceptual flow depicted in Figure~\ref{fig:Approach}.
The process begins in the \emph{Domain Engineering} phase, where we formalize the variability of design patterns. This is the most crucial step and the main contribution of our work. 
We identify the structural and behavioral variation points of each pattern and encode them as UVL feature models. In parallel, we define reusable code generation templates for each pattern. In this context, each design pattern is treated as a reusable asset within the SPL, with its own internal variability explicitly captured in a UVL feature model. This makes it possible to configure and derive pattern implementations in a systematic and automated way.
\begin{itemize}[noitemsep,nolistsep,leftmargin=\parindent]
    \item \textbf{Feature models:} We define the configurable options of each design pattern using UVL. For example, for a specific pattern, we model its structural and behavioral alternatives, and the constraints between them, allowing for fine-grained control over its final implementation. These models serve as formal specifications of the pattern's variability.
    \item \textbf{Implementation artifacts:} For each design pattern, we create language-specific templates (\eg\ written in Swift for iOS) that contain placeholders and Jinja directives based on the features of the UVL model. These templates are the blueprints for the final code, ready to be populated with specific feature selections.
\end{itemize}

In the \emph{Application Engineering} phase, a specific mobile application product is derived by selecting a valid configuration of features for each required design pattern. 
A developer or product manager selects the desired features from the feature model generating a configuration of the feature model (normally stored in a JSON file). 
The final step is the generation of the source code by instantiating the Jinja templates with the provided configuration. As the core engine for this phase, we use a generic variability resolution engine: \emph{UVengine}~\cite{Horcas2025_UVengine}, a tool for resolving the variability of UVL models over Jinja templates. The output is a complete, custom-tailored desing pattern ready to be incorporated in the mobile application. 

In the following section, we provide the details of how to formalize the variability of each design pattern in UVL and how to implement such variability using Jinja templates for Swift code.

\section{Formalizing and Implementing Design Pattern Variability}
\label{sec:Details}
This section focuses on the formalization and systematic implementation of the internal variability of design patterns as configurable assets within the SPL engineering process. First, we describe how to explicitly model the variation points and variants of each pattern using UVL~\cite{Benavides2025_UVL}. Subsequently, we explain how this specification is translated into reusable implementation artifacts through parameterized code templates (using Jinja), which are capable of automatically generating concrete versions of the pattern in response to different application configurations.

\subsection{Overview of Patterns Analyzed}
We present the catalog of design patterns selected for variability modeling and automated code generation. We assign identifiers to each pattern (\eg\ DP1, DP2), in order to facilitate references throughout the following sections.

The patterns chosen represent three classical groups (creational, behavioral, structural) and are among those most frequently used in mobile development. Namely, \emph{Singleton} (DP1), \emph{Strategy} (DP2), \emph{Observer} (DP3), \emph{Adapter} (DP4), and \emph{Factory Method} (DP5). Each pattern exhibits distinct forms of internal variability, both in structural and behavioral aspects. This diversity allows us to generalize the feasibility and scalability of our approach.
For each pattern, we systematically:

\begin{enumerate}[noitemsep,nolistsep,leftmargin=\parindent]
    \item Identify the main variation points (\eg\ instantiation strategy, interface alternatives, cardinality, etc.).
    \item Capture alternatives, constraints, and dependencies in a UVL feature model.
    \item Provide a reusable, parameterized Jinja code template that materializes the selected features.
    \item Analyze the configuration space to determine the expressiveness and boundaries of their variability.
\end{enumerate}
This approach ensures not only a comprehensive exploration of variability (covering both fine-grained options and ``macro'' structural features), but also a uniform methodology for the subsequent sections, where each pattern will be formalized, implemented, and its configuration space evaluated (Sections~4.2--4.6).

By covering a representative sample of patterns and modeling their variability explicitly, we lay the ground for future expansion to other patterns and domains. Table~\ref{tab:patterns} summarizes the selected patterns and their main sources of variability, serving as a roadmap for the remainder of this section.

\begin{table}[ht]
\centering
\caption{Overview of patterns and main sources of variability.}
\label{tab:patterns}
\vspace{-0.4cm}
\scriptsize
\begin{tabular}{lp{1.8cm}p{1.8cm}p{1.8cm}}
\toprule
\textbf{Design Pattern} & \textbf{Structural\newline Variability} & \textbf{Behavioral\newline Variability} & \textbf{Key Constraints} \\
\midrule
DP1: Singleton & Initialization strategy. & Thread safety, access method. & Eager vs lazy, valid combinations. \\

DP2: Strategy & Number and type of strategies. & Dynamic selection mechanism. & Caching, setter/factory selection. \\

DP3: Observer & Observer cardinality. & Sync/async notification. & One-to-many, notification mode. \\

DP4: Adapter & Adaptee binding mechanism. & Interface compatibility. & Inheritance or composition. \\

DP5: Factory Method & Product class alternatives. & Instantiation parameterization. & Subclass responsibility, type safety. \\
\bottomrule
\end{tabular}
\end{table}
\subsection{DP1: Singleton Pattern Variability}
The Singleton pattern is one of the most widely used in mobile application development to ensure the existence of a single instance of a class and to provide a global point of access to it. However, this pattern is not monolithic: it exhibits relevant internal variability that includes aspects such as initialization strategy, access method, and concurrency safety.

\paragraph{\textbf{Variability specification}}
The UVL model in Listing~\ref{uvl:Singleton} captures the main variation points of the Singleton pattern. It defines:
\begin{description}[noitemsep,nolistsep,leftmargin=\parindent]
    \item[\feature{Initialization}:] either \feature{Eager} (the instance is created at class loading time) or \feature{Lazy} (the instance is created on demand).
    \item[\feature{Access Method}:] the instance can be retrieved through a \feature{getInstance()} method or via a static field.
    \item[\feature{Thread Safety}:] synchronization can be added for concurrent scenarios, but it is only applicable in the case of \feature{Lazy} initialization.
    \item[\feature{Class Name}:] defined as a String feature, allowing customization of the generated class name.
\end{description}

\begin{lstlisting}[language=uvl,caption={Feature model of the Singleton pattern.}, label={uvl:Singleton}]
features
    "Singleton Pattern" {abstract}
        mandatory
            Initialization {abstract}
                alternative
                    Eager
                    Lazy
            "Access Method" {abstract}
                alternative
                    "getInstance()"
                    "Static Field"
        optional
            "Thread Safety"
            String "Class Name"
constraints
    Eager => !"Thread Safety"
    "Static Field" => Eager
\end{lstlisting}

Additionally, two constraints are defined to ensure configuration validity: (1) if initialization is \feature{Eager}, the \feature{Thread Safety} option cannot be selected; (2) the \feature{Static Field} access method is only valid in combination with \feature{Eager} initialization.
This model not only formalizes the available options but also prevents inconsistent configurations, ensuring correctness in the generated code.

\begin{lstlisting}[language=jinja,caption={Jinja template excerpt for the Singleton pattern.}, label={jinja:Singleton}]
{# singleton.swift.j2 #}
{% if features.get("Thread Safety") %} import Dispatch {% endif %}
{% set class_name = features.get("Class Name", "Singleton") %}

class {{ class_name }} {
    {% if features.get("Initialization") == "Eager" %}
    static let shared = {{ class_name }}()
    {% elif features.get("Initialization") == "Lazy" %}
    private static var _instance: {{ class_name }}?
    {% endif %}
    
    private init() {} // Private constructor

    {% if features.get("Access Method") == "getInstance()" %}
    static func getInstance() -> {{ class_name }} {
        {% if features.get("Initialization") == "Eager" %}
        return shared
        {% else %}
        // Lazy initialization and Thread Safety options ...
        {% endif %}
    }
    {% elif features.get("Access Method") == "Static Field" %}
    static var instance: {{ class_name }} {
        // Access through static property ...
    }
    {% endif %}
}
\end{lstlisting}

\paragraph{\textbf{Variability implementation}}
Listing~\ref{jinja:Singleton} shows an excerpt of the Jinja template that implements the variability of the Singleton pattern in Swift. The template uses directives to activate or deactivate code fragments depending on the features selected in the UVL model. For instance, the \feature{Thread Safety} option imports the \texttt{Dispatch} library (line 3) and introduces a synchronization queue to protect instance creation. The initialization mechanism is controlled with conditionals: in \feature{Eager} mode, the instance is created statically (line 8), while in \feature{Lazy} mode, it is delayed until the first invocation (line 10). The access method can be implemented either through a \texttt{getInstance()} function (lines 16--22) or through a static property \texttt{instance} (lines 24--26). Finally, the class name is parameterized via the UVL feature \feature{Class Name} (line 4), enabling reuse in different contexts.

\paragraph{\textbf{Variability analysis}}
The configuration space of the Singleton pattern is relatively small but sufficiently expressive to cover the main usage scenarios in mobile applications: 2 initialization options (\feature{Eager} vs. \feature{Lazy}); 2 access methods (\feature{getInstance()} vs. \feature{Static Field}); and the presence or absence of \feature{Thread Safety} (restricted to \feature{Lazy} initialization).
Considering the defined constraints, the model produces a total of four valid configurations\footnote{We exclude the String feature \feature{Class Name} from the analysis since it does not affect the configuration space.}:
\begin{description}[noitemsep,nolistsep,leftmargin=\parindent]
\footnotesize
    \item[Config 1.] \{ \feature{Eager}, \feature{getInstance()} \}
    \item[Config 2.] \{ \feature{Eager}, \feature{Static Field} \}
    \item[Config 3.] \{ \feature{Lazy}, \feature{getInstance()} \}
    \item[Config 4.] \{ \feature{Lazy}, \feature{getInstance()}, \feature{Thread Safety} \}
\end{description}
This analysis confirms that the UVL model captures the essential variability without introducing inconsistent combinations. 
\subsection{DP2: Strategy Pattern Variability}
The Strategy pattern is widely employed in mobile application development as it defines a family of algorithms, encapsulates them, and enables their interchangeability, thereby allowing the algorithm to evolve independently of the clients that employ it. Nevertheless, this pattern is not monolithic: it exhibits significant internal variability that encompasses aspects such as the mechanism for selecting the strategy, the degree of coupling with the context, and performance considerations in resource-constrained environments.

\begin{lstlisting}[language=uvl,caption={Feature model of the Strategy pattern.}, label={uvl:Strategy}]
features
    "Strategy Pattern" {abstract}
        mandatory
            "Strategy Interface" {abstract}
                alternative
                    "Abstract Class"
                    Interface
                    Protocol
            "Strategy Selection" {abstract}
                alternative
                    "Constructor Injection" 
                    "Setter Method"
                    "Factory Method"
            "Context Behavior" {abstract}
                alternative
                    "Delegate to Strategy"
                    "Template Method"
        optional
            "Strategy Validation"
            "Default Strategy"
            "Strategy Caching"
            String "Context Class Name"
            String "Strategy Interface Name"
            Integer "Number of Strategies" cardinality [2..10]
constraints
    "Abstract Class" => !Protocol
    Protocol => Interface
    "Factory Method" => "Strategy Validation"
    "Template Method" => "Abstract Class"
    "Number of Strategies" > 1
\end{lstlisting}

\paragraph{\textbf{Variability specification}} 
The UVL model in Listing~\ref{uvl:Strategy} captures the main variation points of the Strategy pattern. It defines:
\begin{description}[noitemsep,nolistsep,leftmargin=\parindent]
\item[\feature{Strategy Interface}:] The abstraction of the strategy can be realized through an interface (if supported by the target language), a protocol (Swift's interface mechanism), or an abstract class (an inheritance-based approach).

\item[\feature{Strategy Selection}:] The process of selecting and configuring strategies can be performed by means of constructor injection (the strategy is set at creation time), setter method (the strategy can be replaced during execution), or factory method (the strategy is created and selected through a factory).

\item[\feature{Context Behavior}:] The context may either delegate directly to the strategy (simple delegation) or apply the template method (in which the context defines the overall structure of the algorithm while strategies implement specific steps).

\item[\feature{Strategy Validation}:] An optional feature enables validation of the strategy's suitability prior to its execution.

\item[\feature{Default Strategy}:] An optional feature that provides a fallback strategy when no specific strategy is set.

\item[\feature{Strategy Caching}:] An optional feature that enables caching of strategy instances to avoid repeated instantiation, improving performance in scenarios with frequent strategy switches.

\item[\feature{Context Class Name} and \feature{Strategy Interface Name}:] They allow customization of generated class and interface names.

\item[\feature{Number of Strategies}:] An integer feature with cardinality that specifies how many concrete strategy implementations will be generated (range 2--10).
\end{description}
Additionally, several constraints ensure configuration validity: (1) \feature{Abstract Class} approach is incompatible with \feature{Protocol}; (2) \feature{Protocol} requires \feature{Interface} selection; (3) \feature{Factory Method} requires \feature{Strategy Validation} for proper error handling; (4) \feature{Template Method} requires \feature{Abstract Class} approach; (5) and \feature{Number of Strategies} must be greater than 1.

\begin{lstlisting}[language=jinja,caption={Jinja template excerpt for the Strategy pattern.}, label={jinja:Strategy}]
{# strategy.swift.j2 #}
{% set ctx_name = features.get("Context Class Name", "Context") %}
{% set strat_name = features.get("Strategy Interface Name", "Strategy") %}
{% set num_strategies = features.get("Number of Strategies", 2) %}

// Strategy Protocol
{% if features.get("Strategy Interface") == "Protocol" %}
protocol {{ strat_name }} {
    func execute()
}
{% else %}
... // Class-based approach
{% endif %}

// Concrete Strategies ({{ num_strategies }} implementations generated)
{% for i in range(1, num_strategies + 1) %}
class ConcreteStrategy{{ i }}: {{ strat_name }} {
    func execute() { print("Executing strategy {{ i }}") }
}
{% endfor %}

class {{ ctx_name }} {
    private var strategy: {{ strat_name }}
    {% if features.get("Strategy Caching") %}
    private var strategyCache: [String: {{ strat_name }}] = [:]
    {% endif %}
    
    init(strategy: {{ strat_name }}) { self.strategy = strategy }
    
    {% if features.get("Strategy Selection") == "Factory Method" %}
    func setStrategy(byType strategyType: String) {
        // Factory-based instantiation with optional caching
        // Full implementation available at GitHub repository
    }
    {% endif %}
    
    func executeStrategy() { strategy.execute() }
}
\end{lstlisting}

\paragraph{\textbf{Variability implementation}} 
Listing~\ref{jinja:Strategy} shows a Jinja template excerpt for the Strategy pattern in Swift. The template begins by extracting configuration parameters from the UVL features (lines 2--4), setting default values for the context name, strategy interface name, and number of strategies. The \feature{Strategy Interface} selection (lines 7--13) determines whether to generate a Swift protocol or use a class-based approach, though the class-based implementation is abbreviated in this excerpt.

The template generates concrete strategy implementations dynamically (lines 16--26) based on the \feature{Number of Strategies} feature, creating the specified number of \texttt{ConcreteStrategy} classes that conform to the selected interface approach. Each generated strategy includes a simple execution method with pattern-specific logic.
The \texttt{Context} class implementation (lines 29--71) varies based on the selected strategy selection mechanism. Constructor \texttt{Injection} is always supported through the init method (lines 35--37), while optional setter methods are generated when \feature{Setter Method} or \feature{Factory Method} selection is chosen (lines 39--43). The \feature{Factory Method} approach (lines 45--66) includes string-based strategy instantiation with a switch statement that maps strategy names to concrete implementations.
Performance optimization through \feature{Strategy Caching} is conditionally included (lines 31--33, 47--52, 62--64), maintaining a dictionary cache of strategy instances to avoid repeated instantiation.

\paragraph{\textbf{Variability analysis}}
The configuration space of the Strategy pattern is moderately complex, reflecting the diverse architectural needs in mobile application development. The model supports 3 strategy interface approaches, 3 selection mechanisms, 2 context behaviors, and various optional features, combined with a configurable number of strategies (2--10).
Considering the defined constraints, the model produces up to 120 distinct configurations. The constraints significantly reduce the theoretical configuration space by eliminating inconsistent combinations, such as \feature{Abstract Class} with \feature{Protocol} or  \feature{Template Method} without \feature{Abstract Class}.
As examples, we show three representative valid configurations:
\begin{description}[noitemsep,nolistsep,leftmargin=\parindent]
\footnotesize
    \item[Config 1.] \{ \feature{Protocol}, \feature{Interface}, \feature{Constructor Injection}, \feature{Delegate to Strategy}, \feature{3 strategies} \}
    \item[Config 2.] \{ \feature{Protocol}, \feature{Interface}, \feature{Setter Method}, \feature{Delegate to Strategy}, \feature{Strategy Caching}, \feature{5 strategies} \}
    \item[Config 3.] \{ \feature{Abstract Class}, \feature{Factory Method}, \feature{Template Method}, \feature{Strategy Validation}, \feature{Strategy Caching}, \feature{4 strategies} \}    
\end{description}

\subsection{DP3: Observer Pattern Variability}
The Observer pattern is extensively used in mobile application development to establish one-to-many dependencies between objects, allowing multiple observers to be automatically notified when a subject's state changes. In mobile applications, this pattern is particularly valuable for implementing reactive user interfaces, data binding mechanisms, model-view synchronization, and event-driven architectures. It exhibits internal variability such as different notification strategies, observer management approaches, and performance optimizations for resource-constrained mobile environments.

\paragraph{\textbf{Variability specification}}
The UVL model in Listing~\ref{uvl:Observer} defines the main variation points of the Observer pattern:

\begin{description}[noitemsep,nolistsep,leftmargin=\parindent]
\item[\feature{Subject Type}:] It can be implemented using an \feature{Abstract Subject} (inheritance-based approach with common functionality) or \feature{Protocol Subject} (interface-based approach for maximum flexibility).

\item[\feature{Observer Interface}:] The observer abstraction can be defined as a \feature{Protocol} (Swift's interface mechanism) or as an \feature{Abstract Class} (when shared behavior among observers is needed).

\item[\feature{Notification Method}:] The communication mechanism between subject and observers can follow \feature{Push Model} (subject sends data directly to observers), \feature{Pull Model} (observers query the subject for updated data), or \feature{Hybrid Model} (combination of both approaches).

\item[\feature{Observer Management}:] Optional feature that provides sophisticated observer lifecycle management, including registration, deregistration, and observer validation.

\item[\feature{Async Notifications}:] Optional feature enabling asynchronous notification delivery, preventing UI blocking and improving performance in mobile applications.

\item[\feature{Filtered Notifications}:] Optional feature allowing selective notification based on observer interests or event types.

\item[\feature{Observer Priority}:] It enables prioritized notification ordering, ensuring critical observers receive updates first.

\item[\feature{Weak References}:] It implements weak reference patterns to prevent retain cycles and memory leaks in mobile applications.

\item[\feature{Event Types}:] It supports typed events and filtering capabilities.

\item[\feature{Subject}, \feature{Observer}, and \feature{Concrete Observer Name}:] They allow customization of generated class and interface names.

\item[\feature{Number of Observers}:] An integer feature with cardinality specifying how many concrete observer implementations will be generated (range 1-10).
\end{description}

\begin{lstlisting}[language=uvl,caption={Feature model of the Observer pattern.}, label={uvl:Observer}]
features
    "Observer Pattern" {abstract}
        mandatory
            "Subject Type" {abstract}
                alternative
                    "Abstract Subject"
                    "Protocol Subject"
            "Observer Interface" {abstract}
                alternative
                    Protocol
                    "Abstract Class"
            "Notification Method" {abstract}
                alternative
                    "Push Model"
                    "Pull Model"
                    "Hybrid Model"
        optional
            "Observer Management"
            "Async Notifications"
            "Filtered Notifications" 
            "Observer Priority"
            "Weak References"
            "Event Types"
            String "Subject Name"
            String "Observer Name"
            String "Concrete Observer Name"
            Integer "Number of Observers" cardinality [1..10]
constraints
    "Observer Priority" => "Observer Management"
    "Filtered Notifications" => "Event Types"
    "Async Notifications" => "Protocol Subject"
    "Number of Observers" > 0
\end{lstlisting}

Several constraints enforce valid configurations: (1) \feature{Observer Priority} requires \feature{Observer Management} selection; (2) \feature{Filtered Notifications} requires \feature{Event Types} for Efficiency. Without event types, all observers would receive all notifications, which would make the concept of ``filtering'' useless; (3) \feature{Async Notifications} requires \feature{Protocol Subject}; (4) and \feature{The number of observers} must be greater than 0.

\paragraph{\textbf{Variability implementation}}
Listing~\ref{jinja:Observer} shows a Jinja template excerpt for the Observer pattern in Swift. 
The template begins by extracting configuration parameters from the UVL features (lines 2--3), setting default values for the \feature{subject name} and \feature{observer interface name}. The Observer Interface is consistently implemented as a protocol (lines 5--12) with AnyObject conformance to support weak references, a critical requirement for mobile applications to prevent memory leaks.

\begin{lstlisting}[language=jinja,caption={Jinja template excerpt for the Observer pattern.}, label={jinja:Observer}]
{# swift_observer.j2 #}
{% set subject_name = features.get("Subject Name", "Subject") %}
{% set obs_name = features.get("Observer Name", "Observer") %}

protocol {{ obs_name }}: AnyObject {
    var observerID: String { get }
    {% if features.get("Push Model") %}
    func update(data: [String: Any])
    {% elif features.get("Pull Model") %}
    func update(subject: {{ subject_name }})
    {% endif %}
}

class {{ subject_name }} {
    private var observers: [{{ obs_name }}] = []
    
    func attach(_ observer: {{ obs_name }}) {
        observers.append(observer)
    }
    
    func notify({% if features.get("Push Model") %}data: [String: Any]{% endif %}) {
        for observer in observers {
            // Notification method varies: Push/Pull/Hybrid models supported
            // Additional features: async delivery, filtering, priority ordering
        }
    }
}
\end{lstlisting}

The notification method selection drives the core communication mechanism between subject and observers. In \feature{Push Model} mode (lines 7--8, 21, 23--24), the subject actively sends data to observers through the update method with a data parameter. In \feature{Pull Model} mode (lines 9--10, 25--26), observers receive minimal notification and must query the subject directly for updated information, providing more control over data access patterns.

The Subject implementation (lines 14--30) maintains an array of observers and provides basic attachment and notification functionality. The notify method varies its signature based on the selected notification model: \feature{Push Model} includes a data parameter (line 21), while Pull Model requires no additional parameters as observers access the subject directly.
While this excerpt demonstrates the core variability implementation, the complete template includes additional features such as observer management, asynchronous notifications, and filtering capabilities based on the selected optional features from the UVL model.

\paragraph{\textbf{Variability analysis}}
The configuration space of the Observer pattern reflects moderate to high complexity, suitable for the diverse architectural requirements in mobile application development. The model supports 2 subject types, 2 observer interface approaches, 3 notification methods, and 6 optional features, combined with configurable observer counts (1--10).
While the constraints reduce the theoretical configuration space by eliminating architecturally inconsistent combinations, such as asynchronous notifications without protocol-based subjects; the model still produces 648 distinct configurations. 
This configuration space addresses the spectrum of reactive programming needs in mobile development, from simple notification mechanisms to sophisticated event-driven architectures.
Representative valid configurations include:

\begin{description}[noitemsep,nolistsep,leftmargin=\parindent]
\footnotesize
   \item[Config 1.] \{ \feature{Protocol Subject}, \feature{Protocol}, \feature{Push Model}, 2 observers \}. 
   \item[Config 2.] \{ \feature{Protocol Subject}, \feature{Protocol}, \feature{Pull Model}, \feature{Observer Management}, \feature{Weak References}, 5 observers \}. 
   \item[Config 3.] \{ \feature{Protocol Subject}, \feature{Protocol}, \feature{Hybrid Model}, \feature{Async Notifications}, \feature{Filtered Notifications}, \feature{Event Types}, \feature{Observer Priority}, 8 observers \}. 
\end{description}

\subsection{DP4: Adapter Pattern Variability}
The Adapter pattern is a structural design pattern commonly employed in mobile application development to enable interoperability between classes featuring incompatible interfaces. This approach is particularly advantageous within mobile applications for integrating third-party libraries, legacy codebases, external APIs, and platform-specific components that diverge from expected interface specifications. Notably, the Adapter pattern is not monolithic; it exhibits significant internal heterogeneity, encompassing various adaptation strategies, interface compatibility techniques, and data transformation procedures tailored to the diverse requirements encountered in contemporary mobile development scenarios.

\begin{lstlisting}[language=uvl,caption={Feature model of the Adapter pattern.}, label={uvl:Adapter}]
features
    "Adapter Pattern" {abstract}
        mandatory
            "Adapter Type" {abstract}
                alternative
                    "Object Adapter"
                    "Class Adapter"
            "Target Interface" {abstract}
                alternative
                    Protocol
                    "Abstract Class"
            "Adapter Implementation" {abstract}
                alternative
                    "Full Implementation"
                    "Partial Implementation"
        optional
            "Method Mapping"
            "Data Conversion"
            "Error Handling"
            "Multiple Adaptees"
            String "Target Interface Name"
            String "Adapter Class Name"
            String "Adaptee Class Name"
            Integer "Number of Methods" cardinality [1..10]
constraints
    "Class Adapter" => "Abstract Class"
    "Multiple Adaptees" => "Object Adapter"
    "Partial Implementation" => "Error Handling"
    "Number of Methods" > 0
\end{lstlisting}

\paragraph{\textbf{Variability specification}} 
The UVL model in Listing~\ref{uvl:Adapter} captures the main variation points of the Adapter pattern. It defines:

\begin{description}[noitemsep,nolistsep,leftmargin=\parindent]
\item[\feature{Adapter Type}:] The adaptation mechanism can be implemented using \feature{Object Adapter} (composition-based approach that wraps the adaptee) or \feature{Class Adapter} (inheritance-based approach using protocol extensions in Swift).

\item[\feature{Target Interface}:] It can be defined as a \feature{Protocol} (Swift's interface mechanism for maximum flexibility) or \feature{Abstract Class} (when shared implementation is needed across adapters).

\item[\feature{Adapter Implementation}:] The completeness of interface adaptation can be \feature{Full Implementation} (all target methods implemented) or \feature{Partial Implementation} (selective method adaptation with error handling for unimplemented methods).

\item[\feature{Method Mapping}:] Optional feature that provides sophisticated mapping between target and adaptee method signatures, parameter transformations, and naming conventions.

\item[\feature{Data Conversion}:] Optional feature enabling automatic data type conversion between incompatible target and adaptee data types, essential for API integration scenarios.

\item[\feature{Error Handling}:] Optional feature providing robust error handling mechanisms for adaptation failures, type conversion errors, and method invocation issues.

\item[\feature{Multiple Adaptees}:] Optional feature supporting adaptation of multiple incompatible classes through a single adapter interface, useful for aggregating legacy systems.

\item[\feature{Target Interface}, \feature{Adapter Class}, and \feature{Adaptee Class Name}:] String features allowing customization of generated class and interface names for different contexts.

\item[\feature{Number of Methods}:] An integer feature with cardinality [1..10] specifying how many methods will be adapted in the interface.
\end{description}

Additionally, several constraints ensure configuration validity: (1) \feature{Class Adapter} requires \feature{Abstract Class} for inheritance-based adaptation; (2) \feature{Multiple Adaptees} requires \feature{Object Adapter} for composition flexibility; (3) \feature{Partial Implementation} requires \feature{Error Handling} for unimplemented method management; and (4) \feature{Number of Methods} must be greater than 0.

\begin{lstlisting}[language=jinja,caption={Jinja template excerpt for the Adapter pattern.}, label={jinja:Adapter}]
{# adapter.swift.j2 #}
{% set target = features.get("Target Interface Name", "Target") %}
{% set adapter = features.get("Adapter Class Name", "Adapter") %}
{% set adaptee = features.get("Adaptee Class Name", "Adaptee") %}

protocol {{ target }} {
    func request(){% if features.get("Data Conversion") %} -> String{% endif %}
}

class {{ adaptee }} {
    func specificRequest(){% if features.get("Data Conversion") %} -> Int{% endif %} {
        // Full implementation available at GitHub repository
    }
}
// Object Adapter implementation
class {{ adapter }}: {{ target }} { 
    private let adaptee: {{ adaptee }}    
    
    init(adaptee: {{ adaptee }}) { self.adaptee = adaptee }    
    
    func request(){% if features.get("Data Conversion") %} -> String{% endif %} {
        {% if features.get("Data Conversion") %}
        let result = adaptee.specificRequest()
        return "Converted: \(result)"
        {% else %}
        adaptee.specificRequest()
        {% endif %}
    }
}
\end{lstlisting}

\paragraph{\textbf{Variability implementation}}
Listing~\ref{jinja:Adapter} shows an excerpt of the Jinja template that implements the variability of the Adapter pattern in Swift. The template uses conditional directives to generate different structural approaches based on the selected features from the UVL model. The template begins by extracting configuration parameters (lines 2--4), setting default values for the \feature{Target Interface Name}, \feature{Adapter Class Name}, and \feature{Adaptee Class Name}. The target interface is consistently implemented as a protocol (lines 6--8), with the \feature{Data Conversion} feature determining whether methods return specific types or void. The adaptee class represents the legacy component being adapted (lines 9--16), with \feature{Data Conversion} controlling return type compatibility. The core adaptation logic switches between \feature{Object Adapter} (lines 18--31) using composition with a private adaptee instance and initialization, versus \feature{Class Adapter} (lines 32--41) using Swift protocol extensions for inheritance-based adaptation. The \texttt{request} method implementation varies based on \feature{Data Conversion}, either performing type conversion from \texttt{Int} to \texttt{String} (lines 26--27, 35--36) or simple method delegation (lines 29, 38). While this excerpt demonstrates the fundamental adaptation mechanisms, the complete template includes additional features such as method mapping, error handling, and multiple adaptee support based on selected optional features.

\paragraph{\textbf{Variability analysis}} 
The configuration space of the Adapter pattern reflects moderate complexity, addressing diverse integration challenges in mobile application development. The model supports 2 adapter types, 2 target interface approaches, 2 implementation completeness levels, and 4 optional features, combined with configurable method counts (1--10).

Considering the defined constraints, the model produces up to 120 distinct valid configurations that takes care of architecturally inconsistent combinations, such as \feature{Class Adapter} without \feature{Abstract Class} or \feature{Multiple Adaptees}
without \feature{Object Adapter} flexibility.
Representative valid configurations include:

\begin{description}[noitemsep,nolistsep,leftmargin=\parindent]
\footnotesize
    \item[Config 1.] \{ \feature{Object Adapter}, \feature{Protocol}, \feature{Full Implementation}, \feature{Data Conversion}, 3 methods \}
    \item[Config 2.] \{ \feature{Class Adapter}, \feature{Abstract Class}, \feature{Partial Implementation}, \feature{Error Handling}, \feature{Method Mapping}, 5 methods \}
    \item[Config 3.] \{ \feature{Object Adapter}, \feature{Protocol}, \feature{Full Implementation}, \feature{Multiple Adaptees}, \feature{Data Conversion}, \feature{Error Handling}, 8 methods \}
 \end{description}

\subsection{DP5: Factory Method Pattern Variability}
The Factory Method pattern occupies a central position in mobile software development when the objective is to delegate object instantiation to subclasses, thereby enhancing the architectural flexibility and extensibility of the system. However, its application presents diverse alternatives and differentiated decisions, each of which can have significant consequences for the final design.

\begin{lstlisting}[language=uvl,caption={Feature model of the Factory Method pattern.}, label={uvl:FactoryMethod}]
features
    "Factory Method Pattern" {abstract}
        mandatory
            "Creator Type" {abstract}
                alternative
                    "Abstract Creator"
                    "Concrete Creator"
            "Product Interface" {abstract}
                alternative
                    Protocol
                    "Abstract Class"
                    "Base Class"
            "Factory Method" {abstract}
                alternative
                    "Abstract Method"
                    "Default Implementation"
        optional
            "Product Registration"
            "Parameter Passing"
            "Error Handling"
            "Generic Support"
            "Lazy Loading"
            String "Creator Class Name"
            String "Product Interface"
            String "Factory Method Name"
            Integer "Number of Products" cardinality [2..10]
constraints
    "Abstract Method" => "Abstract Creator"
    "Default Implementation" => "Concrete Creator"
    "Product Registration" => "Generic Support"
    "Number of Products" > 1
\end{lstlisting}

\paragraph{\textbf{Variability specification}}
The UVL model in Listing~\ref{uvl:FactoryMethod} captures the main variation points of the Factory Method pattern:
\begin{description}[noitemsep,nolistsep,leftmargin=\parindent]
\item[\feature{Creator Type}:] The pattern accommodates both an \feature{Abstract Creator} that specifies the factory interface, and a \feature{Concrete Creator} that provides the factory implementation.

\item[\feature{Product Interface}:] It is manifested as a \feature{Protocol}, \feature{Abstract Class}, or \feature{Base Class}, enabling a spectrum of polymorphism and type abstraction suitable for different software requirements.

\item[\feature{Factory Method}:] There are two main options available for implementation: an \feature{Abstract Method} (which requires subclasses to provide the instantiation logic) or a \feature{Default Implementation} (the base class provides a default creation method).

\item[\feature{Optional Features}:] The model encompasses additional extensions such as \feature{Product Registration} (enabling runtime registration of available products), \feature{Parameter Passing} (the factory method can accept arguments that influence instance creation), \feature{Error Handling}, \feature{Generic Support}, and \feature{Lazy Loading}.

\item[\feature{String and Integer Attributes}:] Configuration is further customized through \feature{Creator Class Name}, \feature{Product Interface Name}, and \feature{Factory Method Name} (all string attributes). The number of products is set via the integer attribute \feature{Number of Products}, with a minimum cardinality of two.
\end{description}

\begin{lstlisting}[language=jinja,caption={Template excerpt for the Factory Method pattern.}, label={jinja:FactoryMethod}]
{# factory_method.swift.j2 #}
{% set creator = features.get("Creator Class Name", "Creator") %}
{% set product = features.get("Product Interface", "Product") %}

protocol {{ product }} {
    func operation(){% if features.get("Parameter Passing") %} -> String{% endif %}
}

class ConcreteProduct: {{ product }} {
    func operation(){% if features.get("Parameter Passing") %} -> String{% endif %} {
        print("ConcreteProduct: operation")
        {% if features.get("Parameter Passing") %}
        return "result"
        {% endif %}
    }
}

{% if features.get("Abstract Creator") %}
class {{ creator }} {
    func createProduct() -> {{ product }} {
        fatalError("Override required")
    }
}
{% else %}
class {{ creator }} {
    func createProduct({% if features.get("Parameter Passing") %}type: String{% endif %}) -> {{ product }} {
        return ConcreteProduct()
    }
}{% endif %}
\end{lstlisting}

\paragraph{\textbf{Variability implementation}}
Listing~\ref{jinja:FactoryMethod} shows an excerpt of the Jinja template that implements the variability of the Factory Method pattern in Swift. The template uses directives to activate or deactivate code fragments depending on the features selected in the UVL model. The template begins by extracting configuration parameters from the UVL features (lines 2--3), setting default values for the \feature{Creator Class Name} and \feature{Product Interface Name}. The product interface is consistently implemented as a protocol (lines 5--7), with the \feature{Parameter Passing} feature determining whether methods include return types or additional parameters. The core factory method implementation switches between \feature{Abstract Creator} (lines 18--23) using \texttt{fatalError("Override required")} to enforce subclass implementation, versus \feature{Concrete Creator} (lines 24--29) providing a default implementation that returns a concrete product instance. The \texttt{createProduct} method varies its signature based on \feature{Parameter Passing}, either accepting type arguments (line 26) or requiring no parameters for simple instantiation.

\paragraph{\textbf{Variability analysis}}
The configuration space of the Factory Method pattern reflects moderate to high complexity, addressing diverse object creation challenges in mobile application development. The model supports 2 creator types, 3 product interface approaches, 2 factory method implementations, and 5 optional features, combined with configurable product counts (2--10).

Considering the defined constraints, the model produces 288 valid configurations that consider architecturally inconsistent combinations, such as avoiding \feature{Abstract Method} without \feature{Abstract Creator} or \feature{Product Registration} without \feature{Generic Support} flexibility.
Representative valid configurations include:

\begin{description}[noitemsep,nolistsep,leftmargin=\parindent]
\footnotesize
    \item[Config 1.] \{ \feature{Abstract Creator}, \feature{Protocol}, \feature{Abstract Method}, \feature{Parameter Passing}, 3 products \}
    \item[Config 2.] \{ \feature{Concrete Creator}, \feature{Base Class}, \feature{Default Implementation}, \feature{Error Handling}, \feature{Lazy Loading}, 5 products \} 
    \item[Config 3.] \{ \feature{Abstract Creator}, \feature{Protocol}, \feature{Abstract Method}, \feature{Product Registration}, \feature{Generic Support}, \feature{Parameter Passing}, \feature{Error Handling}, 8 products \} 
\end{description}

\section{Evaluation}
\label{sec:Evaluation}
We make our UVL models and Jinja templates publicly available and conducted experiments to analyse the variability complexity and the configuration space of the design patterns.
In addition, we assess the development effort involved by comparing manual and generated implementations, illustrating this with a case study on the Singleton pattern.

\subsection{Dataset Availability and Open Science}
Following open science best practices, our UVL models dataset as well as the Jinja templates and all associated resources are publicly available online in \emph{GitHub}:

\begin{itemize}[noitemsep,nolistsep,leftmargin=\parindent]
    \item Website and instructions: \textbf{\url{https://trran.github.io/UVL2Pat/}} 
    \item GitHub: \textbf{\url{https://github.com/trran/UVL2Pat}}
\end{itemize}

\subsection{Experimentation Setup}
We use the tools \emph{flamapy}~\cite{Galindo2023_Flama} and \emph{FM Fact Label}~\cite{Horcas2025_FMFactLabel}, that are part of the UVL ecosystem~\cite{Galindo2024_OpenScienceTutorial}, to analyze the complexity and configuration space of the feature models specifed in UVL. Those are online web-based tools, and thus, not additional setup is required.

\subsection{Results and Discussion}
The complexity of a feature model can be characterized by its number of features and its number of valid products (configurations)~\cite{Bagheri2011_StructuralMetrics,clements_2001}. The goal is to analyze the complexity of the UVL models and the configuration space of the design patterns.

\begin{figure}[t]
    \centering
    \includegraphics[width=\linewidth,height=4cm]{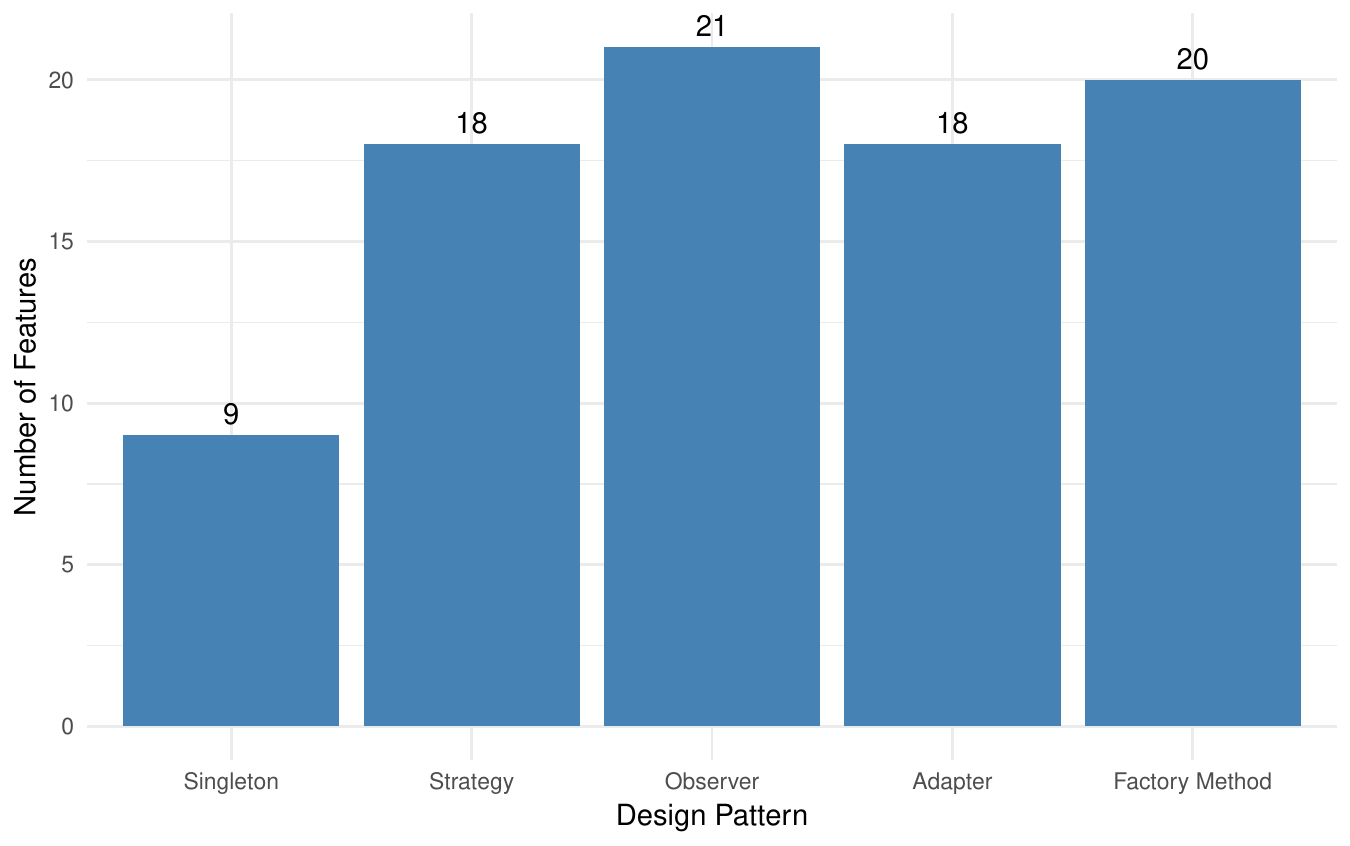}
    \vspace{-0.8cm}
    \caption{Comparison of patterns in terms of features.}
    \label{fig:FeaturesComparison}
    \vspace{-0.4cm}
\end{figure}

\begin{figure}[t]
    \centering
    \includegraphics[width=\linewidth,height=4cm]{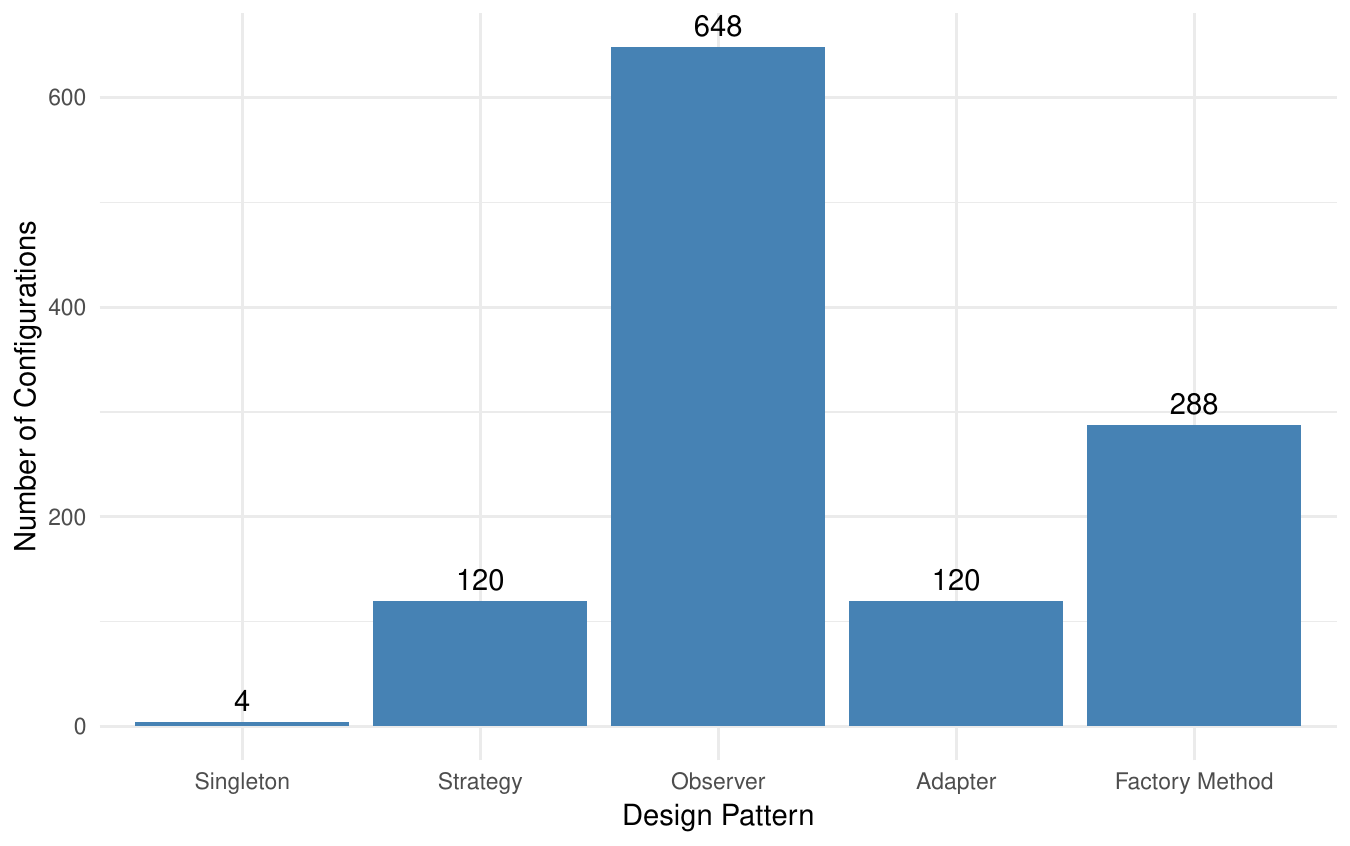}
    \vspace{-0.8cm}
    \caption{Comparison of the patterns' configuration space.}
    \label{fig:ConfigurationsComparison}
    \vspace{-0.4cm}
\end{figure}

Figure~\ref{fig:FeaturesComparison} summarizes the patterns according to the number of features captured in their UVL models, while Figure~\ref{fig:ConfigurationsComparison} compares them with respect to the size of their configuration spaces.
From Figure~\ref{fig:FeaturesComparison}, we observe that the variability of the patterns is almost evenly distributed. Except for simpler patterns such as Singleton, which exhibit nine features, the others patterns involve a richer feature set (from 18 to 21 different features), as they encapsulate more variation points related to roles, interactions, and optional extensions. 
Despite the number of features is similar along the patterns, they present different type of features and variation points for each pattern (as shown in Section~\ref{sec:Details}). This suggests that different patterns introduce different levels of modeling effort: some can be formalized with minimal feature hierarchies, while others require a more fine-grained specification to capture their inherent flexibility.

Figure~\ref{fig:ConfigurationsComparison} further highlights the impact of variability on the configuration space. The Singleton pattern results in only four valid configurations due to its restricted options and constraints. The Factory Method and Observer patterns, on the other hand, generate a significantly larger configuration space (288 and 648 configurations respectively), which reflects their combinatorial variability when multiple roles and communication mechanisms are considered. The Strategy and Adapter patterns occupy an intermediate position: although their feature models are not as complex as Factory Method or Observer, their configuration spaces are still non-trivial due to the inclusion of optional features and multiple mapping options.

These findings lead to several insights. First, the formalization of design patterns as feature models makes their variability explicit and comparable, which is often overlooked in traditional pattern catalogs. Second, the analysis of configuration spaces reveals the potential challenges in automated derivation: while some patterns can be generated with almost no configuration overhead, others may require careful selection of features to avoid overwhelming the developer with unnecessary options. Finally, the results confirm that our approach scales from simple to more complex patterns, showing that UVL and template-based generation are suitable for capturing and exploiting variability across the full spectrum of design patterns.

\paragraph{\textbf{Effort Analysis: Singleton Pattern case study}}
To further illustrate the benefits of our approach, we compared a manually implemented Singleton, inspired by a real-world chat service, with the automatically generated version produced by our UVL + Jinja pipeline. The manual implementation consisted of multiple classes (\eg\ \texttt{FriendsChatService}, \texttt{Message}, and two view controllers) and approximately 130 lines of code. In contrast, the generated version focused exclusively on the structural essence of the pattern, with only 55 lines of code and built-in thread safety.
Table~\ref{tab:singleton-comparison} summarizes the comparison. The results show that our approach reduces boilerplate code by more than 50\%, lowers cyclomatic complexity, and introduces configuration options (\eg\ lazy initialization, thread safety) not present in the manual implementation. While domain-specific logic still requires manual coding, the proposed approach accelerates prototyping and ensures architectural consistency.

\begin{table}[ht]
\vspace{-0.2cm}
\caption{Development effort comparison of Singleton pattern.}
\label{tab:singleton-comparison}
\centering
\vspace{-0.4cm}
\scriptsize
\begin{tabular}{lcc}
\toprule
\textbf{Metric} & \textbf{Manual (Chat)} & \textbf{Generated (UVL+Jinja)} \\
\midrule
Lines of Code (LOC) & $\sim$130 & $\sim$55 \\

Number of Classes & 6 & 1 \\

Cyclomatic Complexity & $\geq$ 10 & $\leq$ 3 \\

Estimated Dev. Time & 3--4 h & 15--20 min \\

Configurability & Fixed & Parametrizable \\

Reusability & Limited to case & Generic \\
\bottomrule
\end{tabular}%
\vspace{-0.6cm}
\end{table}

\subsection{Threats to Validity}
This work presents certain limitations that may affect the evaluation validity. First, the evaluation has focused on a reduced set of patterns and exclusively on one language (Swift). It remains necessary to broaden the empirical validation to include more design patterns such as Decorator, Proxy, or State, among others; as well as other programming languages used in mobile applications such as Kotlin, Java, or Dart. Another threat is the lack of a comparison between the effort and benefits relative to the automated generation of the patterns' variants and the manual development of those variants.


\section{Related Work} 
\label{sec:RelatedWork}
This section discusses prior research related to our work on modeling and exploiting design pattern variability for mobile application generation. We organize the discussion into two main areas: approaches that combine SPL with design patterns, and existing generative techniques in software development.

\paragraph{\textbf{Design Patterns and SPLs}}

The intersection of SPLs and design patterns has been explored primarily through patterns as variability implementation mechanisms in SPLs ~\cite{Apel2013_FOSPL,Czarnecki2000_GenerativeProgramming}. The Strategy pattern implements behavioral variability, Decorator enables dynamic feature addition, Template Method varies algorithms through inheritance, and Observer facilitates feature addition/removal. However, our work addresses a different challenge: the internal variability within patterns themselves, which has received limited attention ~\cite{Seidl2017_GenerativeSPLDesignPatterns}.

Our work, however, tackles a different problem: the variability within the patterns themselves. The intrinsic configurable nature of design patterns has been largely overlooked, with variations often being implicit in standard catalogs~\cite{Gamma1995_DesignPatterns,Freeman2004_HeadFirstDesignPatterns,Shvets2018_DiveIntoDesignPatterns}. A notable exception is the work by Seidl et al.~\cite{Seidl2017_GenerativeSPLDesignPatterns}, who introduced the concept of \emph{variability-aware design patterns} to bridge the gap between SPLs and design patterns. Their approach uses \emph{role models} to capture a design pattern and its connection to an SPL feature model, allowing for systematic application and generation of pattern-based artifacts.
Other related efforts include those that combine feature models with architectural patterns. Font et al.~\cite{Font2015} proposed a method for building SPLs from conceptualized model patterns. Degueule et al.~\cite{Degueule2017} developed a tooling solution that integrates architectural pattern catalogs with variability models in CVL (the obsolete \emph{Common Variability Language})~\cite{Haugen2008_CVL} to automatically synthesize architectural variants. While these works are highly relevant, they often rely on predefined pattern models and explicit metamodels. Our approach, by using a modern and extensible language such as UVL and template-based code generation, provides a flexible and scalable method to formalize and exploit design pattern variability. Moreover, UVL enables the automated analysis of the configuration space of the design patterns by using the tool support of the UVL ecosystem~\cite{Galindo2024_OpenScienceSPL} such as \emph{flamapy}~\cite{Galindo2023_Flama}, \emph{flamapyIDE}~\cite{Sebastian2025_FlamapyIDE}, \emph{FeatureIDE}~\cite{Sundermann2021_UVLFeatureIDE}, \emph{FM Fact Label}~\cite{Horcas2025_FMFactLabel}, and \emph{UVengine}~\cite{Horcas2025_UVengine}.

\paragraph{\textbf{Generative Programming and Mobile Application Development}}
The field of Generative Programming~\cite{Czarnecki2000_GenerativeProgramming} focuses on creating software from high-level specifications. The implementation of SPLs and the automatic generation of products relies on different approaches. These methods can be broadly categorized into \emph{composition-based} and \emph{annotation-based} techniques~\cite{Apel2013_FOSPL,Horcas2022_CombiningMultipleGranularity}. Compositional approaches, such as those used in \emph{Feature-Oriented Software Development} (FOSD)~\cite{Apel2013_FOSPL}, \emph{Aspect-Oriented Programming} (AOP)~\cite{Kiczales1997}, or \emph{Delta-Oriented Programming} (DOP)~\cite{Schaefer2010}, among others, rely on composing code fragments or assets to build a complete product. In contrast, annotation-based techniques use annotations or directives within a code base to mark variation points, which are then processed by a generator to create a specific product variant (\eg\ preprocessors~\cite{Hunsen2016_CPP} or configuration parameters).
Although the literature often recommends the use of compositional over annotation-based approaches due to their advantages in managing complexity and reusability~\cite{Apel2013_FOSPL}, the industry has adopted annotative approaches because of their simplicity and lower entry barrier~\cite{Horcas2022_CombiningMultipleGranularity,Hunsen2016_CPP}.

A widely adopted and effective technique within this field is \emph{template-based code generation}~\cite{Syriani2018_SMS_TemplateCodeGeneration}. This approach enriches a base language with special directives to automatically produce customized content. While sharing similarities with annotation-based techniques, our approach leverages the power of the Jinja templates, which offer significant advantages over simpler annotative-based approaches. Unlike tools that are limited to basic \texttt{\#ifdef} annotations~\cite{Hunsen2016_CPP} or simple parameter substitution~\cite{Cortinas2022_spljsengine}, Jinja provides a rich set of directives for managing complex variability. This includes advanced control structures such as \emph{if/elif/else}, \emph{for-loops}, \emph{macros}, and \emph{blocks}, which enable sophisticated logic for substitution and replacement of code elements.
Furthermore, Jinja's language-independent nature makes it a highly versatile tool, capable of being used to generate code for any text-based language, including mobile applications. To effectively bridge the gap between our high-level variability models and the code generation process, our approach relies on \emph{UVengine}~\cite{Horcas2025_UVengine}. This universal variability resolution engine is specifically designed to resolve UVL models over Jinja templates. Its utility has been demonstrated in practical scenarios across diverse domains and artifact types, such as: \emph{Visualization pipelines} to generate data visualization charts with configurable visual components and datasets~\cite{RomeroOrganvidez2024_DataVisualizationSPL,Horcas2022_DataVisualizationVariability}; \emph{Kubernetes configurations} to derive deployment descriptors for microservices and containers, adapting resource constraints based on the Kubernetes feature model~\cite{Horcas2025_KubernetesFM}; and \emph{Data migration processes} to generate reusable transformation scripts in data migration between heterogeneous content management systems~\cite{RomeroOrganvidez2024_DataTransformationVariability}. However, to the best of our knowledge, our work is the first to apply this specific generative approach to model and exploit the internal variability of design patterns for the purpose of generating customized mobile application code.

While automatic code generation is increasingly popular, especially with the rise of large language models (LLMs)~\cite{Juyong2024_SurveyLLMsCodeGeneration}, the resulting code often lacks architectural quality and adherence to fundamental design principles~\cite{Mulla2024_ChoosingBestArchitecture,Nguyen2018_DeepLearningUIDesignPatterns}. Our work addresses this gap by providing a structured, model-driven approach that ensures the systematic application of design patterns, thereby promoting the generation of well-architected mobile applications. The specific context of mobile application development presents a unique challenge, as it requires a high degree of adaptability to different platforms (\eg\ iOS, Android) and evolving user interfaces, while maintaining performance and code quality. By focusing on Swift and the generation of customizable design patterns, our work provides a concrete solution to the specific challenges faced by mobile developers who require scalable and maintainable solutions across different platforms and user expectations.

\section{Conclusions and Future Work}
\label{sec:Conclusions}

In this work, we have addressed the following research question: \emph{How can the variability of design patterns be modeled and exploited to generate customized and architecturally robust mobile applications?}
Our principal contribution is the formalization of the internal variability of classic design patterns through the Universal Variability Language (UVL), combined with their integration into Jinja templates for the automatic generation of Swift code. This approach enables design patterns to be treated as configurable assets within an SPL approach, opening the door to mobile applications that are more flexible, maintainable, and adaptable.

The analysis of the configuration space demonstrates that the variability of each pattern can be systematically and consistently captured, thus avoiding invalid configurations and ensuring correct implementations. This evidences the viability of bridging model-driven engineering with practical mobile code generation, guaranteeing architectural quality even in the context of AI-assisted development or rapid code generation tools.

For future work, we plan to expand our research in several directions: (1) expanding our catalog of supported patterns; (2) incorporating quality and maintainability metrics in the evaluation; (3) exploring integration with generative AI tools; and (4) applying this methodology to industrial-scale mobile development case studies.




\begin{acks}
This work is supported by Mescyt, Universidad San Jorge, ITIS/Universidad de Málaga, and Sergio Jiménez González (Hotel Las Galias), to whom we are deeply grateful.
\end{acks}

\bibliographystyle{ACM-Reference-Format}
\bibliography{references}

@String{Computer = "{IEEE} Computer" }

@String{Springer = "Springer-Verlag" }

@Book{Felfernig2024_FMsAIDriven,
  author    = {Alexander Felfernig and Andreas A. Falkner and David Benavides},
  publisher = {Springer},
  title     = {Feature Models - AI-Driven Design, Analysis and Applications},
  year      = {2024},
  isbn      = {978-3-031-61873-4},
  series    = {Springer Briefs in Computer Science},
  bibsource = {dblp computer science bibliography, https://dblp.org},
  doi       = {10.1007/978-3-031-61874-1},
  url       = {https://doi.org/10.1007/978-3-031-61874-1},
}

@article{Benavides2025_UVL,
title = {{UVL}: Feature modelling with the Universal Variability Language},
journal = {J. Syst. Softw.},
volume = {225},
year = {2025},
issn = {0164-1212},
doi = {https://doi.org/10.1016/j.jss.2024.112326},
author = {David Benavides and Chico Sundermann and Kevin Feichtinger and José A. Galindo and Rick Rabiser and Thomas Thüm},
}

@Book{Apel2013_FOSPL,
  author    = {Sven Apel and Don S. Batory and Christian K{\"{a}}stner and Gunter Saake},
  publisher = {Springer},
  title     = {Feature-Oriented Software Product Lines - Concepts and Implementation},
  year      = {2013},
  isbn      = {978-3-642-37520-0},
  bibsource = {dblp computer science bibliography, https://dblp.org},
  doi       = {10.1007/978-3-642-37521-7},
  url       = {https://doi.org/10.1007/978-3-642-37521-7},
}

@inproceedings{Galindo2023_Flama,
  author       = {Jos{\'{e}} A. Galindo and Jos{\'{e}} Miguel Horcas and Alexander Felfernig and David Fern{\'{a}}ndez{-}Amor{\'{o}}s and David Benavides},
  title        = {{FLAMA:} {A} collaborative effort to build a new framework for the
                  automated analysis of feature models},
  booktitle    = {27th {ACM} International Systems and Software Product Line Conference ({SPLC})},
  pages        = {16--19},
  publisher    = {{ACM}},
  volume    = {B},
  year         = {2023},
  url          = {https://doi.org/10.1145/3579028.3609008},
  doi          = {10.1145/3579028.3609008},
  bibsource    = {dblp computer science bibliography, https://dblp.org}
}

@inproceedings{Font2015,
  author = {Font, Jaime and Arcega, Lorena and {\O}ystein Haugen and Cetina, Carlos},
  title = {Building software product lines from conceptualized model patterns},
  booktitle = {17th International Software Product Line Conference (SPLC 2015)},
  year = {2015},
  pages = {46--55}
}

@inproceedings{Degueule2017,
author = {Degueule, Thomas and Filho, Joao Bosco Ferreira and Barais, Olivier and Acher, Mathieu and Le Noir, J\'{e}r\^{o}me and Madel\'{e}nat, S\'{e}bastien and Gailliard, Gr\'{e}gory and Burlot, Godefroy and Constant, Olivier},
title = {Tooling support for variability and architectural patterns in systems engineering},
year = {2015},
isbn = {9781450336130},
publisher = {ACM},
url = {https://doi.org/10.1145/2791060.2791097},
doi = {10.1145/2791060.2791097},
booktitle = {19th International Conference on Software Product Line ({SPLC})},
pages = {361–364},
numpages = {4},
}

@book{Gamma1995_DesignPatterns,
  title={Design patterns: elements of reusable object-oriented software},
  author={Gamma, Erich and Helm, Richard and Johnson, Ralph and Vlissides, John},
  year={1995},
  publisher={Pearson GmbH},
}

@article{Punchoojit2017_UsabilityStudiesSLR,
author = {Punchoojit, Lumpapun and Hongwarittorrn, Nuttanont},
title = {Usability Studies on Mobile User Interface Design Patterns: A Systematic Literature Review},
journal = {Advances in Human-Computer Interaction},
volume = {2017},
number = {1},
pages = {6787504},
doi = {https://doi.org/10.1155/2017/6787504},
url = {https://onlinelibrary.wiley.com/doi/abs/10.1155/2017/6787504},
eprint = {https://onlinelibrary.wiley.com/doi/pdf/10.1155/2017/6787504},
year = {2017}
}

@inproceedings{Nguyen2018_DeepLearningUIDesignPatterns,
author = {Nguyen, Tam The and Vu, Phong Minh and Pham, Hung Viet and Nguyen, Tung Thanh},
title = {Deep learning UI design patterns of mobile apps},
year = {2018},
isbn = {9781450356626},
publisher = {ACM},
url = {https://doi.org/10.1145/3183399.3183422},
doi = {10.1145/3183399.3183422},
booktitle = {40th International Conference on Software Engineering ({ICSE}): New Ideas and Emerging Results},
pages = {65–68},
numpages = {4},
}

@article{Zaina2022_GuidelinesUIDesignPatterns,
title = {Preventing accessibility barriers: Guidelines for using user interface design patterns in mobile applications},
journal = {Journal of Systems and Software},
volume = {186},
pages = {111213},
year = {2022},
issn = {0164-1212},
doi = {https://doi.org/10.1016/j.jss.2021.111213},
url = {https://www.sciencedirect.com/science/article/pii/S0164121221002831},
author = {Luciana A.M. Zaina and Renata P.M. Fortes and Vitor Casadei and Leornardo Seiji Nozaki and Débora Maria Barroso Paiva},
}

@Article{DaSilva2022_MobileUIDesignPatternsSMS,
AUTHOR = {da Silva, Leonardo Filipe and Parreira Junior, Paulo Afonso and Freire, André Pimenta},
TITLE = {Mobile User Interaction Design Patterns: A Systematic Mapping Study},
JOURNAL = {Information},
VOLUME = {13},
YEAR = {2022},
NUMBER = {5},
ARTICLE-NUMBER = {236},
URL = {https://www.mdpi.com/2078-2489/13/5/236},
ISSN = {2078-2489},
DOI = {10.3390/info13050236}
}

@book{Orlova2024_FlutterDesignPatterns,
  title={Flutter Design Patterns and Best Practices},
  author={Orlova, Daria and Kadah, Esra},
  year={2024},
  publisher ={Packt Publishing}
}

@article{Mulla2024_ChoosingBestArchitecture,
  title={Choosing the Best Architecture for Mobile Applications},
  author={Mulla, Farooq},
  journal={ResearchGate Publication, December},
  year={2024}
}

@article{Juyong2024_SurveyLLMsCodeGeneration,
  author       = {Juyong Jiang and
                  Fan Wang and
                  Jiasi Shen and
                  Sungju Kim and
                  Sunghun Kim},
  title        = {A Survey on Large Language Models for Code Generation},
  journal      = {CoRR},
  volume       = {abs/2406.00515},
  year         = {2024},
  url          = {https://doi.org/10.48550/arXiv.2406.00515},
  doi          = {10.48550/ARXIV.2406.00515},
  eprinttype    = {arXiv},
  eprint       = {2406.00515},
  bibsource    = {dblp computer science bibliography, https://dblp.org}
}

@TechReport{Kang1990_FODA,
  author      = {Kang, Kyo C and Cohen, Sholom G and Hess, James A and Novak, William E and Peterson, A Spencer},
  institution = {Carnegie-Mellon Univ Pittsburgh Pa Software Engineering Inst},
  title       = {Feature-oriented domain analysis ({FODA}) feasibility study},
  year        = {1990},
  url         = {https://resources.sei.cmu.edu/library/asset-view.cfm?assetid=11231},
}

@Book{Pohl2005_SPLEngineering,
  author    = {Klaus Pohl and G{\"{u}}nter B{\"{o}}ckle and Frank van der Linden},
  publisher = {Springer},
  title     = {Software Product Line Engineering - Foundations, Principles, and Techniques},
  year      = {2005},
  isbn      = {978-3-540-24372-4},
  bibsource = {dblp computer science bibliography, https://dblp.org},
  doi       = {10.1007/3-540-28901-1},
  url       = {https://doi.org/10.1007/3-540-28901-1},
}

@Book{Czarnecki2000_GenerativeProgramming,
  author    = {Krzysztof Czarnecki and Ulrich W. Eisenecker},
  publisher = {Addison-Wesley},
  title     = {Generative programming - methods, tools and applications},
  year      = {2000},
  isbn      = {978-0-201-30977-5},
  bibsource = {dblp computer science bibliography, https://dblp.org},
}

@article{Seidl2017_GenerativeSPLDesignPatterns,
title = {Generative software product line development using variability-aware design patterns},
journal = {Computer Languages, Systems \& Structures},
volume = {48},
pages = {89-111},
year = {2017},
issn = {1477-8424},
doi = {https://doi.org/10.1016/j.cl.2016.08.006},
url = {https://www.sciencedirect.com/science/article/pii/S1477842415300609},
author = {Christoph Seidl and Sven Schuster and Ina Schaefer}
}

@book{Shvets2018_DiveIntoDesignPatterns,
  title={Dive into design patterns},
  author={Shvets, Alexander},
  publisher={Refactoring. Guru},
  year={2018}
}

@Article{Czarnecki2005_CardinalityBasedFM,
  author    = {Krzysztof Czarnecki and Simon Helsen and Ulrich W. Eisenecker},
  journal   = {Softw. Process. Improv. Pract.},
  title     = {Formalizing cardinality-based feature models and their specialization},
  year      = {2005},
  number    = {1},
  pages     = {7--29},
  volume    = {10},
  bibsource = {dblp computer science bibliography, https://dblp.org},
  doi       = {10.1002/spip.213},
  url       = {https://doi.org/10.1002/spip.213},
}

@InProceedings{Sundermann2021_UVL,
  author    = {Chico Sundermann and Kevin Feichtinger and Dominik Engelhardt and Rick Rabiser and Thomas Th{\"{u}}m},
  booktitle = {25th {ACM} International Systems and Software Product Line Conference ({SPLC})},
  title     = {Yet another textual variability language?: a community effort towards a unified language},
  year      = {2021},
  pages     = {136--147},
  volume    = {A},
  doi       = {10.1145/3461001.3471145},
  url       = {https://doi.org/10.1145/3461001.3471145},
}

@InProceedings{Sundermann2023_UVLParserExtensions,
  author    = {Chico Sundermann and Stefan Vill and Thomas Th{\"{u}}m and Kevin Feichtinger and Prankur Agarwal and Rick Rabiser and Jos{\'{e}} A. Galindo and David Benavides},
  booktitle = {27th {ACM} International Systems and Software Product Line Conference ({SPLC})},
  title     = {{UVLParser}: Extending {UVL} with Language Levels and Conversion Strategies},
  year      = {2023},
  address   = {Tokyo, Japan},
  month     = sep,
  pages     = {39--42},
  publisher = {{ACM}},
  bibsource = {dblp computer science bibliography, https://dblp.org},
  doi       = {10.1145/3579028.3609013},
  url       = {https://doi.org/10.1145/3579028.3609013},
}

@article{Syriani2018_SMS_TemplateCodeGeneration,
  author    = {Eugene Syriani and
               Lechanceux Luhunu and
               Houari A. Sahraoui},
  title     = {Systematic mapping study of template-based code generation},
  journal   = {Comput. Lang. Syst. Struct.},
  volume    = {52},
  pages     = {43--62},
  year      = {2018},
  url       = {https://doi.org/10.1016/j.cl.2017.11.003},
  doi       = {10.1016/j.cl.2017.11.003},
  bibsource = {dblp computer science bibliography, https://dblp.org}
}

@inproceedings{Horcas2025_UVengine,
  author       = {Jos{\'{e}} Miguel Horcas and M{\'{o}}nica Pinto and Lidia Fuentes},
  title        = {\emph{UVengine}: A Universal Variability Resolution Engine for Feature Models Using Template-Based Artifacts},
  booktitle    = {29th {ACM} International Systems and Software Product Line Conference ({SPLC})},
  publisher    = {{ACM}},
   volume    = {B},
  year         = {2025},
   address   = {A Coruña, Spain},
  month     = sep,
  url          = {https://doi.org/10.1145/3748269.3748486},
  doi          = {10.1145/3748269.3748486}
}

@book{Freeman2004_HeadFirstDesignPatterns,
  asin = {0596007124},
  author = {Freeman, Eric and Freeman, Elisabeth and Bates, Bert and Sierra, Kathy},
  dewey = {005.1},
  ean = {9780596007126},
  edition = 1,
  isbn = {0596007124},
  publisher = {O'Reilly Media},
  title = {Head First Design Patterns},
  year = 2004
}

@inproceedings{Galindo2024_OpenScienceSPL,
  author       = {Jos{\'{e}} A. Galindo and
                  David Romero{-}Organvidez and
                  Megha Bhushan and
                  Jos{\'{e}} Miguel Horcas Aguilera and
                  David Benavides},
  title        = {Open Science principles in software product lines: The case of the {UVL} ecosystem},
  booktitle    = {28th {ACM} International Systems and Software Product Line Conference ({SPLC})},
  volume       = {A},
  pages        = {223},
  year         = {2024},
  url          = {https://doi.org/10.1145/3646548.3674550},
  doi          = {10.1145/3646548.3674550},
  bibsource    = {dblp computer science bibliography, https://dblp.org}
}

@inproceedings{Sebastian2025_FlamapyIDE,
  author       = {Francisco Sebastian Benitez and
                  Jos{\'{e}} A. Galindo and
                  David Romero{-}Organvidez and
                  David Benavides},
  title        = {{UVL} web-based editing and analysis with flamapy.ide},
  booktitle    = {19th International Working Conference on Variability Modelling of Software-Intensive Systems ({VaMoS})},
  pages        = {121--125},
  year         = {2025},
  url          = {https://doi.org/10.1145/3715340.3715436},
  doi          = {10.1145/3715340.3715436},
  bibsource    = {dblp computer science bibliography, https://dblp.org}
}

@article{Horcas2025_FMFactLabel,
  author       = {Jos{\'{e}} Miguel Horcas and
                  Jos{\'{e}} A. Galindo and
                  Lidia Fuentes and
                  David Benavides},
  title        = {{FM} fact label},
  journal      = {Sci. Comput. Program.},
  volume       = {240},
  year         = {2025},
  url          = {https://doi.org/10.1016/j.scico.2024.103214},
  doi          = {10.1016/J.SCICO.2024.103214},
  bibsource    = {dblp computer science bibliography, https://dblp.org}
}

@InProceedings{Sundermann2021_UVLFeatureIDE,
  author    = {Chico Sundermann and Tobias He{\ss} and Dominik Engelhardt and Rahel Arens and Johannes Herschel and Kevin Jedelhauser and Benedikt Jutz and Sebastian Krieter and Ina Schaefer},
  booktitle = {25th {ACM} International Systems and Software Product Line Conference ({SPLC})},
  title     = {Integration of {UVL} in FeatureIDE},
  year      = {2021},
  volume    = {B},
  doi       = {10.1145/3461002.3473940},
  url       = {https://doi.org/10.1145/3461002.3473940},
}

@article{Horcas2022_CombiningMultipleGranularity,
  author       = {Jos{\'{e}} Miguel Horcas and Alejandro Corti{\~{n}}as and Lidia Fuentes and Miguel R. Luaces},
  title        = {Combining multiple granularity variability in a software product line approach for web engineering},
  journal      = {Inf. Softw. Technol.},
  volume       = {148},
  pages        = {106910},
  year         = {2022},
  url          = {https://doi.org/10.1016/j.infsof.2022.106910},
  doi          = {10.1016/J.INFSOF.2022.106910},
  bibsource    = {dblp computer science bibliography, https://dblp.org}
}

@InProceedings{Kiczales1997,
  author       = {Kiczales, Gregor and Lamping, John and Mendhekar, Anurag and Maeda, Chris and Lopes, Cristina and Loingtier, Jean-Marc and Irwin, John},
  title        = {Aspect-oriented programming},
  booktitle    = {European conference on object-oriented programming},
  year         = {1997},
  pages        = {220--242},
  organization = {Springer},
}

@InProceedings{Schaefer2010,
  author    = {Schaefer, Ina and Bettini, Lorenzo and Damiani, Ferruccio and Tanzarella, Nico},
  title     = {Delta-oriented Programming of Software Product Lines},
  booktitle = {14th International Conference on Software Product Lines ({SPLC}): Going Beyond},
  year      = {2010},
  pages     = {77--91},
  publisher = {Springer-Verlag},
  acmid     = {1885647},
  isbn      = {3-642-15578-2, 978-3-642-15578-9},
  numpages  = {15},
  url       = {http://dl.acm.org/citation.cfm?id=1885639.1885647},
}

@article{Hunsen2016_CPP,
  author    = {Claus Hunsen and
               Bo Zhang and
               Janet Siegmund and
               Christian K{\"{a}}stner and
               Olaf Le{\ss}enich and
               Martin Becker and
               Sven Apel},
  title     = {Preprocessor-based variability in open-source and industrial software
               systems: An empirical study},
  journal   = {Empirical Software Engineering},
  volume    = {21},
  number    = {2},
  pages     = {449--482},
  year      = {2016},
  url       = {https://doi.org/10.1007/s10664-015-9360-1},
  doi       = {10.1007/s10664-015-9360-1},
  bibsource = {dblp computer science bibliography, https://dblp.org}
}

@inproceedings{Cortinas2022_spljsengine,
  author       = {Alejandro Corti{\~{n}}as and
                  Miguel R. Luaces and
                  Oscar Pedreira},
  title        = {\emph{spl-js-engine}: a JavaScript tool to implement software product lines},
  booktitle    = {26th {ACM} International Systems and Software Product Line Conference ({SPLC})},
  volume       = {B},
  pages        = {66--69},
  year         = {2022},
  url          = {https://doi.org/10.1145/3503229.3547035},
  doi          = {10.1145/3503229.3547035},
  bibsource    = {dblp computer science bibliography, https://dblp.org}
}

@article{RomeroOrganvidez2024_DataVisualizationSPL,
  author       = {David Romero{-}Organvidez and Jos{\'{e}} Miguel Horcas and Jos{\'{e}} A. Galindo and David Benavides},
  title        = {Data visualization guidance using a software product line approach},
  journal      = {J. Syst. Softw.},
  volume       = {213},
  pages        = {112029},
  year         = {2024},
  url          = {https://doi.org/10.1016/j.jss.2024.112029},
  doi          = {10.1016/J.JSS.2024.112029},
  bibsource    = {dblp computer science bibliography, https://dblp.org}
}

@inproceedings{Horcas2022_DataVisualizationVariability,
  author       = {Jos{\'{e}} Miguel Horcas and Jos{\'{e}} A. Galindo and David Benavides},
  title        = {Variability in data visualization: a software product line approach},
   booktitle    = {26th {ACM} International Systems and Software Product Line Conference ({SPLC})},
  pages        = {55--66},
   volume    = {A},
  year         = {2022},
  url          = {https://doi.org/10.1145/3546932.3546993},
  doi          = {10.1145/3546932.3546993},
  bibsource    = {dblp computer science bibliography, https://dblp.org}
}

@inproceedings{Horcas2025_KubernetesFM,
  author       = {Jos{\'{e}} Miguel Horcas and
                  Mercedes Amor Pinilla and
                  Lidia Fuentes},
  title        = {The Kubernetes variability model},
  booktitle    = {19th International Working Conference on Variability
                  Modelling of Software-Intensive Systems ({VaMoS})},
  year         = {2025},
  url          = {https://doi.org/10.1145/3715340.3715440},
  doi          = {10.1145/3715340.3715440},
  bibsource    = {dblp computer science bibliography, https://dblp.org}
}

@inproceedings{RomeroOrganvidez2024_DataTransformationVariability,
  author       = {David Romero{-}Organvidez and David Benavides and Jos{\'{e}} Miguel Horcas and Mar{\'{\i}}a Teresa G{\'{o}}mez{-}L{\'{o}}pez},
  title        = {Variability in data transformation: towards data migration product lines},
  booktitle    = {18th International Working Conference on Variability Modelling of Software-Intensive Systems ({VaMoS})},
  year         = {2024},
  url          = {https://doi.org/10.1145/3634713.3634724},
  doi          = {10.1145/3634713.3634724},
  bibsource    = {dblp computer science bibliography, https://dblp.org}
}

@InProceedings{Haugen2008_CVL,
  author    = {{\O}ystein Haugen and Birger M{\o}ller{-}Pedersen and Jon Oldevik and G{\o}ran K. Olsen and Andreas Svendsen},
  booktitle = {12th International Conference on Software Product Lines ({SPLC})},
  title     = {Adding Standardized Variability to Domain Specific Languages},
  year      = {2008},
  address   = {Limerick, Ireland},
  month     = sep,
  pages     = {139--148},
  publisher = {{IEEE} Computer Society},
  bibsource = {dblp computer science bibliography, https://dblp.org},
  doi       = {10.1109/SPLC.2008.25},
  url       = {https://doi.org/10.1109/SPLC.2008.25},
}

@InProceedings{Galindo2024_OpenScienceTutorial,
  author       = {José A. Galindo and David Romero-Organvidez and Megha Bhushan and José Miguel Horcas Aguilera and David Benavides},
  booktitle    = {28th {ACM} International Systems and Software Product Line Conference ({SPLC})},
  title        = {{Open Science} principles in software product lines: The case of the {UVL} ecosystem},
  year         = {2024},
  month        = sep,
  pages        = {223},
  volume       = {A},
  bibsource    = {dblp computer science bibliography, https://dblp.org},
  doi          = {10.1145/3646548.3674550},
  url          = {https://doi.org/10.1145/3646548.3674550},
  track        = {Tutorial}
}

@article{Bagheri2011_StructuralMetrics,
author = {Bagheri, Ebrahim and Gasevic, Dragan},
year = {2011},
month = {09},
pages = {579-612},
title = {Assessing the maintainability of software product line feature models using structural metrics},
volume = {19},
journal = {Software Quality Journal},
doi = {10.1007/s11219-010-9127-2}
}

@book{clements_2001,
author={Clements, Paul and Northrop, Linda},
title={Software Product Lines: Practices and Patterns},
month={Aug},
year={2001},
howpublished={Carnegie Mellon University, Software Engineering Institute's Digital Library},
}



\end{document}